\documentclass[aps,prl,twocolumn,superscriptaddress,longbibliography]{revtex4-2}
\usepackage{graphicx}
\usepackage{amssymb,amsmath}
\usepackage{bm}
\usepackage{dcolumn}
\usepackage{float}
\usepackage[OT1]{fontenc} 
\usepackage{url}
\usepackage{mathrsfs}
\usepackage{slashed,comment}
\usepackage{color}
\usepackage{verbatim}
\usepackage{enumitem}
\usepackage{soul,physics}
\usepackage[driverfallback=dvipdfm]{hyperref}
\hypersetup{pdfpagemode=FullScreen,colorlinks=true,breaklinks,urlcolor=blue,linkcolor=blue,citecolor=blue}

\usepackage{subfigure}
\usepackage{amssymb}
\usepackage{bm}
\usepackage{graphicx}
\usepackage{amsmath,amssymb}

\begin{document}
  
  \title{Augmenting Quantum Neural Networks with Contractive Layers}

  \author{Bolin Huang}
  \affiliation{State Key Laboratory of Surface Physics \& Department of Physics \& Center for Field Theory and Particle Physics, Fudan University, Shanghai, 200438, China}

  \author{Pengfei Zhang}
  \thanks{PengfeiZhang.physics@gmail.com}
  \affiliation{State Key Laboratory of Surface Physics \& Department of Physics \& Center for Field Theory and Particle Physics, Fudan University, Shanghai, 200438, China}
  \affiliation{Hefei National Laboratory, Hefei 230088, China}

  \date{\today}

  \begin{abstract}
  Classical neural networks combine diverse layers with complementary roles to achieve rich expressive capabilities across complex learning tasks. Rapid advances in quantum science and technology have motivated growing interest in their quantum counterparts, quantum neural networks, raising the question of how their expressivity can be enhanced through the design of new quantum layers. Of particular interest are layers that enrich the accessible function space while introducing only a small number of additional trainable parameters. In this Letter, we introduce the \textit{contractive layer} as a promising building block for quantum neural networks. This layer comprises all-to-all couplings governed by \textit{a single} trainable parameter, enabling efficient control of quantum information flow while maintaining experimental relevance to Rydberg atom arrays. Using observable learning as a benchmark, we show that contractive layers enhance expressivity more efficiently than simply increasing the depth of conventional brick-wall circuits. This advantage stems from a systematic increase in the accessible target-space dimension per trainable parameter. We further demonstrate similar improvements in image recognition tasks. These results highlight contractive layers as a practical route to expressive quantum neural networks on near-term quantum devices.
  \end{abstract}
    
  \maketitle

  \emph{ \color{blue}Introduction.--} Classical neural networks have become a cornerstone of modern machine learning~\cite{LeCun:2015pmr}, enabling advances in tasks ranging from image recognition~\cite{NIPS2012_c399862d,Simonyan2014VeryDC,He_2016_CVPR,Dosovitskiy2020AnII} to natural language processing~\cite{NIPS2014_5a18e133,Vaswani2017AttentionIA,Devlin2019BERTPO}. They combine diverse layers with complementary roles to learn expressive representations and generalize across complex learning tasks. For example, batch normalization layers facilitate training by normalizing intermediate activations~\cite{Ioffe2015BatchNA}, while dropout layers regularize training by randomly deactivating units~\cite{Hinton2012ImprovingNN,Srivastava2014DropoutAS}. This architectural diversity highlights the importance of layer design in shaping the capabilities of neural networks. Rapid advances in quantum science and technology have motivated growing interest in quantum neural networks, in which parameterized quantum circuits operate on input quantum data~\cite{Benedetti:2017ehk,Beach:2019zte,Du:2018epv,Benedetti:2019cxw,Liu:2018obn,Funcke:2020vkw,Huggins:2018kof,Cong:2019uyt,Torrontegui:2019cqu,Jager:2025lpd,Benedetti:2019inj,Schuld:2018ahn,McClean:2018jps,Grant:2018oml,Verdon:2018zkf,Farhi:2018nhu,Zeng:2018rox,Mitarai:2018voy,Beer:2019dxy,Zhang:2024ktz,Zhang:2024xjo,Chen:2025dkp,Hoffmann:2026bhe,Scarato:2026wbh,Mauser:2025cho,Yuan:2026mil,Lakhdar-Hamina:2025kpu}. Their expressive capabilities depend on the structure and composition of these quantum operations, motivating the exploration of quantum layers with distinct and complementary functions~\cite{Sim:2019yyv,Schuld:2020enb,Wu:2021qlx}. Meanwhile, near-term quantum devices impose practical constraints on circuit depth and computational resources~\cite{Preskill:2018jim,Bharti:2021zez}. Consequently, a central challenge is to expand the accessible function space while introducing only a small number of additional trainable parameters.

  Applying quantum circuits to input quantum data can be viewed as a quantum dynamical process. Consequently, guiding principles for designing quantum neural networks often stem from the theory of quantum dynamics. For example, studies of random quantum circuits have motivated the introduction of random layers into quantum neural networks~\cite{PhysRevLett.132.010602}. Related work has also explored architectures with multiple parallel instances~\cite{Choi:2026uup,Zhao:2026jtw,Nguyen2025QMoEAQ,Beckmann:2026dsf,Friedrich:2024qnu,Incudini:2023zyk,Tolotti:2025qjm,Coyle:2024ylr,Park:2026vlh,Rhrissorrakrai2026QuantumEM}. Meanwhile, Ref.~\cite{Shen:2019uap} establishes a connection between quantum machine learning and information scrambling, motivating the derivation of bounds on training error in terms of quantum scrambling~\cite{Garcia:2021bin}. These connections suggest that circuit structures with strong quantum information processing capabilities may provide useful quantum layers. One such structure has recently been introduced in classical shadow tomography~\cite{Huang:2020aa}: a \textit{contractive unitary} built from all-to-all couplings that reduces operator size, thereby improving measurement efficiency~\cite{Wu:2024eno,Nie:2026qcg}. However, whether contractive unitaries can also serve as building blocks for expressive quantum neural networks remains unexplored.

  \begin{figure}[t]
    \centering
    \includegraphics[width=0.9\linewidth]{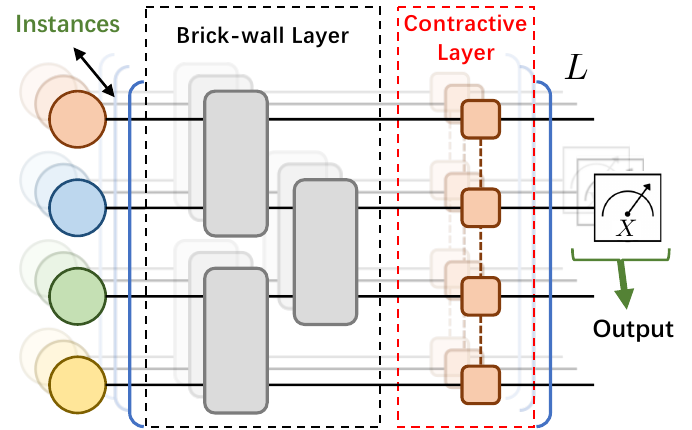}
    \caption{Schematic illustration of a quantum neural network architecture augmented with contractive layers. Each instance processes the input quantum data using a parameterized quantum circuit incorporating brick-wall layers and contractive layers of the form $\hat{U}_{\text{CT}}=e^{i\alpha \prod_i\hat{Z}_i}$, built from all-to-all couplings and controlled by a single trainable parameter $\alpha$. Measurement outputs from the parallel instances are combined through a trainable classical function to produce the final prediction. }
    \label{fig:schematics}
  \end{figure}

  In this Letter, we address this issue by generalizing the contractive unitary to trainable contractive layers and demonstrating their effectiveness in enhancing the expressivity of quantum neural networks with multiple parallel instances (see FIG.~\ref{fig:schematics} for an illustration). Using observable learning as a concrete example, we show that this new architecture outperforms networks composed solely of conventional brick-wall layers, even though each contractive layer introduces only a single trainable parameter. Our analysis links the improved learning performance to the growth of the accessible target-space dimension, showing that adding contractive layers makes more effective use of trainable parameters than simply increasing circuit depth. We further demonstrate their utility in classical learning tasks through image recognition. Our results establish contractive layers as a promising ingredient for constructing expressive quantum neural networks on near-term quantum devices and suggest potential applications in quantum algorithms more broadly.

  \emph{ \color{blue}Architecture.--} We begin with a detailed description of our quantum neural network architecture incorporating contractive layers. The goal of quantum machine learning is to learn a target function $y(\psi)$ of input data encoded in a quantum state $\ket{\psi}$. In the setup of multi-instance quantum neural networks, $R$ copies of the input state are prepared, where $R$ denotes the number of instances, and each copy undergoes a parameterized quantum circuit specified by an independent set of parameters. We consider two different building blocks of the quantum circuits, as illustrated in FIG.~\ref{fig:schematics}. The first is the brick-wall quantum circuit, which consists of alternating layers of parameterized two-qubit gates acting on neighboring qubits. A single brick-wall layer yields
  \begin{equation}
  \hat{U}_{\text{BW}}(\bm{\theta})=\prod_{i\in \text{even}}\hat{U}_{i,i+1}(\bm{\theta}_i)\prod_{i\in \text{odd}}\hat{U}_{i,i+1}(\bm{\theta}_i).
  \end{equation}
  Here, $\hat{U}_{i,i+1}(\bm{\theta}_i)$ denotes a generic two-qubit gate acting on the $i$th and $(i+1)$th qubits, parameterized by 15 real parameters $\bm{\theta}_i$ corresponding to the SU(4) group. We then collect all these parameters into $\bm{\theta}$ for conciseness.

  The second building block, which is the main focus of this work, is the \textit{contractive layer}. It consists of all-to-all couplings with a single trainable rotation angle $\alpha$:
  \begin{equation}
  \hat{U}_{\text{CT}}(\alpha)
  =\exp\Big(i\alpha \prod_i\hat{Z}_i\Big).
  \end{equation}
  Here, $\hat{Z}_i$ denotes the Pauli-$Z$ operator acting on the $i$th qubit. When $\alpha=\pi/4$, this layer reproduces an equivalent form of the contractive unitary introduced in Ref.~\cite{Wu:2024eno}, which rapidly reduces the size of complex operators in a nearly optimal manner~\cite{Nie:2026qcg}. Making the rotation angle $\alpha$ trainable allows the layer to adapt its contractive strength to the learning task. This coupling is also well suited to reconfigurable Rydberg atom arrays, where flexible atom rearrangement facilitates efficient implementation of the gate using a single ancilla~\cite{Jaksch:2000cus,Weimer_2010,Isenhower:2009rsz,Ebadi:2022oxd,Bluvstein:2021jsq,Manetsch:2024lwl,Finkelstein:2024zgp,Evered:2026jrf,White:2026jwt,Miles:2026fbp}. On other quantum platforms, the gate can also be implemented using a circuit whose depth scales linearly with the number of qubits~\cite{Wu:2024eno}.

  The total evolution for a single instance $r$ is determined by how the two building blocks are combined. Here, we focus on two representative scenarios: (1) the circuit contains only brick-wall layers, giving $\hat{U}^{r}=\hat{U}_{\text{BW}}(\bm{\theta}^{r,L})\cdots\hat{U}_{\text{BW}}(\bm{\theta}^{r,1})$; (2) the circuit alternates between brick-wall and contractive layers, yielding $\hat{U}^{r}=\hat{U}_{\text{CT}}({\alpha}^{r,L})\hat{U}_{\text{BW}}(\bm{\theta}^{r,L})\cdots\hat{U}_{\text{CT}}({\alpha}^{r,1})\hat{U}_{\text{BW}}(\bm{\theta}^{r,1})$. Here, $L$ denotes the number of repeated circuit units, and each layer has an independent set of trainable parameters. After the unitary evolution, we measure a local observable $\hat{X}_{\frac{N+1}{2}}$ on the central qubit, where $N$ is the total number of qubits and is assumed to be odd. This yields the expectation value 
  \begin{equation}
  m^{r}(\psi)=\bra{\psi}(\hat{U}^{r})^{\dagger}\hat{X}_{\frac{N+1}{2}}\hat{U}^{r}\ket{\psi}.
  \end{equation}
  Finally, we combine the measurement outcomes from different instances through a simple classical function to obtain the final prediction $\hat{y}(\psi)$. The choice of this function depends on the specific task and will be specified later. Our central focus is to analyze how the loss after training decreases as the numbers of layers and instances increase, comparing architectures with and without contractive layers.

  \begin{figure}[t]
    \centering
    \includegraphics[width=0.98\linewidth]{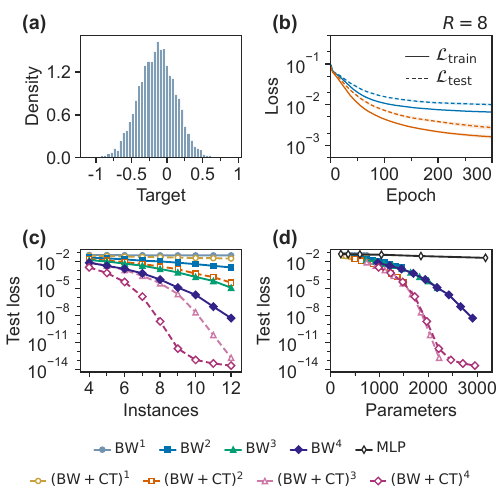}
    \caption{Numerical results for the observable learning task. (a) Distribution of the target values $y(\psi)$ in the training set. (b) Training and test loss as a function of training epochs, with the number of instances fixed at $R=8$. (c-d) Test loss after training as a function of the number of instances and the number of trainable parameters, respectively. Different curves correspond to architectures with or without contractive layers and $L\in\{1,2,3,4\}$. }
    \label{fig:ob}
  \end{figure}

  \emph{ \color{blue}Observable Learning.--} We first consider observable learning, where the target is the expectation value of a generic operator $\hat{O}$ in the input state, namely $y(\psi)=\bra{\psi}\hat{O}\ket{\psi}$. In our numerical simulation, both the operator $\hat{O}$ and the input quantum states are randomly generated using the procedure detailed in the Supplementary Material~\cite{SM}. As shown in FIG.~\ref{fig:ob}(a), the target values in the training set exhibit a broad distribution. Since the target is a linear function of the input density matrix $\ket{\psi}\bra{\psi}$, we use a linear classical readout, and the quantum neural network prediction reads
  \begin{equation}
  \hat{y}(\psi)=\sum_{r=1}^R\left(a^r m^r(\psi)+b^r\right)=\bra{\psi}\hat{M}(\bm{\xi})\ket{\psi},
  \end{equation}
  where the effective measurement operator is given by $\hat{M}(\bm{\xi})=\sum_{r=1}^R\left[a^r(\hat{U}^r)^{\dagger}\hat{X}_{\frac{N+1}{2}}\hat{U}^r+b^r\hat{I}\right]$, with $\bm{\xi}$ collecting all trainable parameters, including $\bm{\theta}^{r,\tau}$, ${\alpha}^{r,\tau}$, $a^r$, and $b^r$. The loss function is given by $\mathcal{L}=\frac{1}{|B|}\sum_{s\in B}(\hat{y}(\psi_s)-y(\psi_s))^2$, where the sum runs over the training set $B$. The parameters are optimized using the standard Adam algorithm with mini-batches~\cite{Kingma2014AdamAM}.

  The numerical results for $N=5$ are presented in FIG.~\ref{fig:ob} for various architectures with or without contractive layers and $L\in\{1,2,3,4\}$. First, both the training and test losses decrease substantially as the number of training epochs increases, as shown in FIG.~\ref{fig:ob}(b) for $R=8$ and $L=2$. The architecture with contractive layers achieves significantly lower losses than the architecture consisting solely of brick-wall layers. To further demonstrate the generality of this observation, we examine the test loss after training for different numbers of instances $R$ and circuit depths $L$ in FIG.~\ref{fig:ob}(c). The results reveal a consistent improvement: circuits with contractive layers at depth $L$ achieve test losses comparable to those of circuits without contractive layers at depth $L+1$ for $L\geq 2$. This is counterintuitive from a naive comparison of parameter counts, since each brick-wall layer contains 60 trainable parameters, whereas each contractive layer contains only one. To illustrate this difference more clearly, we replot the test loss as a function of the number of trainable parameters in FIG.~\ref{fig:ob}(d). The curves within each architecture class exhibit a clear data collapse, with contractive layers consistently yielding lower losses across the numbers of instances and depths considered. We also compare the quantum neural networks with a classical multilayer perceptron (MLP) and find that the quantum models achieve substantially lower test losses.

  \begin{figure}[t]
    \centering
    \includegraphics[width=0.98\linewidth]{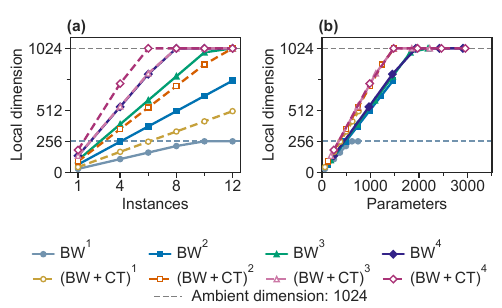}
    \caption{Numerical results for the growth of the local dimension (defined in the main text) as a function of the number of instances or trainable parameters. Different curves correspond to architectures with or without contractive layers and $L\in\{1,2,3,4\}$. The results exhibit a data collapse similar to that observed for the test loss in FIG.~\ref{fig:ob}(c-d). }
    \label{fig:ob2}
  \end{figure}

  \emph{ \color{blue}Local Dimension.--} To understand this systematic improvement, we analyze the local dimension of the operator space explored by the effective measurement operator under small variations in the trainable parameters. This dimension characterizes the local expressivity of the parameterized quantum circuits. Related approaches quantify local expressivity through the rank of the state Jacobian or the quantum Fisher information matrix, thereby identifying independent parameter directions and redundancies~\cite{Funcke:2020vkw,Haug:2021bca,Yao:2025fwz}. Here, we consider the effective measurement operator rather than the output state, following an observable-space perspective also used in Jacobian-based dimension analysis~\cite{Haferkamp2021LinearGO,Ohno:2026sve,Cha:2026byk}. We first expand the measurement operator as $\hat{M}(\bm{\xi})=\sum_P h_P(\bm{\xi})P$ in the complete basis of Pauli string operators $\{P\}$. The coefficients $h_P(\bm{\xi})$ are real and depend smoothly on the trainable parameters. To first order, a small parameter variation $\delta\bm{\xi}$ induces a change $\delta h_P=\sum_\mu J_{P\mu}(\bm{\xi})\delta\xi_\mu$, where $J_{P\mu}(\bm{\xi})=\partial h_P(\bm{\xi})/\partial\xi_\mu$ is the Jacobian matrix. We define the local dimension as $d_{\mathrm{loc}}(\bm{\xi})=\operatorname{rank}J(\bm{\xi})$, which counts the linearly independent directions in operator space accessible through infinitesimal parameter variations. 
  
  Although the local dimension can be analyzed theoretically for random training parameters with $L=1$, as discussed in the Supplementary Material~\cite{SM}, here we focus on numerical results near the parameter configurations obtained after training. The results are presented in FIG.~\ref{fig:ob2}(a-b). FIG.~\ref{fig:ob2}(a) shows that the local dimension increases linearly for small numbers of instances, where the contributions $m^r(\psi)$ from different instances are approximately independent. For a single brick-wall layer, the circuit light cone restricts the support of the measurement operator to a four-qubit subsystem, so the accessible operator space has a maximum dimension of 256. A similar restriction occurs with one additional contractive layer, as discussed in the Supplementary Material~\cite{SM}. In all other cases, the local dimension eventually saturates at 1024, the dimension of the operator space of the entire system. The efficiency of contractive layers at depth $L$ is reflected in a slope comparable to that of the architecture without contractive layers at depth $L+1$ for $L\geq 2$. An upper bound for the slope is derived in the Supplementary Material~\cite{SM}, which matches the numerical results. Furthermore, FIG.~\ref{fig:ob2}(b) shows that the local-dimension curves for different values of $R$ collapse within each architecture class when plotted against the number of trainable parameters. 

  These results are consistent with the behavior of the test loss, supporting the interpretation that contractive layers improve learning performance by expanding the accessible operator space more efficiently. To assess the generality of this observation, we also evaluate the local rank at randomly sampled parameter configurations and find the same enhancement, demonstrating its robustness beyond the configurations obtained through training. Since the local rank is defined independently of the target observable, it provides a task-independent characterization of the local expressivity of quantum neural network architectures. These findings suggest that the benefits of contractive layers may extend to a broad range of quantum machine learning tasks.

  \begin{figure}[t]
    \centering
    \includegraphics[width=0.98\linewidth]{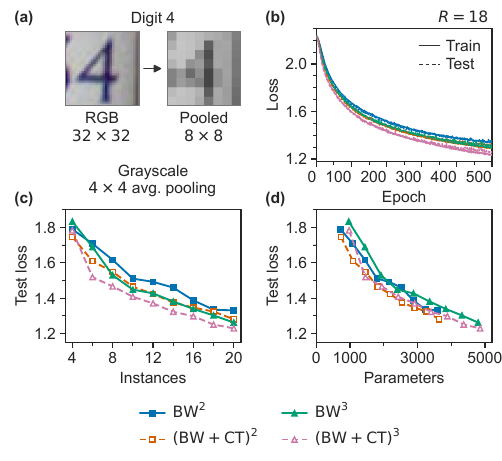}
    \caption{Numerical results for the image recognition task. (a) A representative image from the dataset and the corresponding input data after grayscale conversion and pooling. (b) Training and test losses as a function of training epochs, with the number of instances fixed at $R=18$. (c-d) Test loss after training as a function of the number of instances and the number of trainable parameters, respectively. Different curves correspond to architectures with or without contractive layers and $L\in\{2,3\}$.}
    \label{fig:image}
  \end{figure}

  \emph{ \color{blue}Image Recognition.--} To demonstrate the benefits of contractive layers across a broader range of learning tasks, we next turn to the practical problem of image recognition. We use the Street View House Numbers (SVHN) dataset, in which each image has a resolution of $32\times32$ pixels and is labeled with a digit $y\in\{0,\ldots,9\}$~\cite{Netzer2011ReadingDI}. The goal is to predict the correct digit label. To encode the images into quantum states, we first convert them to grayscale and apply pooling to reduce their dimensionality. The resulting image, illustrated in FIG.~\ref{fig:image}(a), contains $8\times8$ pixels, each represented by a real number. We then flatten the image into a $64$-dimensional real vector and use its components as the real and imaginary parts of a $32$-dimensional complex vector. After normalization, this vector defines the wave function of the input quantum state $\ket{\psi}$. For this ten-class recognition problem, we use a trainable classical readout to map the measurement outcomes from the $R$ instances to ten class probabilities $p_k$, with $k\in\{0,1,\ldots,9\}$, satisfying $\sum_k p_k=1$ and $p_k>0$, as detailed in the Supplementary Material~\cite{SM}. Each $p_k$ represents the predicted probability that the input belongs to class $k$. The loss function is chosen as the cross entropy, $\mathcal{L}=-\frac{1}{|B|}\sum_{s\in B}\log p_{y(\psi_s)}(\psi_s)$, where $B$ denotes the training set and $y(\psi_s)$ is the true digit label of sample $s$.

  The numerical results are presented in FIG.~\ref{fig:image}(b-d). For $R=18$, both the training and test losses decrease steadily as training proceeds, with contractive layers yielding lower losses at each depth considered [FIG.~\ref{fig:image}(b)]. This advantage persists across different numbers of instances, as shown in FIG.~\ref{fig:image}(c). As in observable learning, the architecture with contractive layers at $L=2$ achieves performance comparable to that of the purely brick-wall architecture at $L=3$, despite requiring significantly fewer trainable parameters. The comparison in FIG.~\ref{fig:image}(d) further demonstrates this parameter efficiency: architectures with contractive layers generally achieve lower test losses for the same number of trainable parameters, although the data collapse is less pronounced than in observable learning. These results demonstrate that the benefits of contractive layers extend beyond observable learning to image recognition, offering a more parameter-efficient way to extract task-relevant information from encoded classical data.

  \emph{ \color{blue}Discussion.--} In this Letter, we have introduced trainable contractive layers as fundamental building blocks for quantum neural networks. When combined with conventional brick-wall layers, they improve learning performance with extremely small parameter overhead. Our numerical results demonstrate this advantage in both observable learning and image recognition, including comparisons with deeper brick-wall circuits. The analysis of the local dimension provides a geometric understanding of this improvement: contractive layers enable more efficient exploration of the accessible operator space, while the observed data collapse demonstrates the robustness of this advantage across different numbers of instances and circuit depths. These results highlight contractive layers as a practical route to expressive quantum neural networks on near-term quantum devices, particularly reconfigurable Rydberg atom arrays, which support all-to-all connectivity.

  Several directions remain open. First, extending the local-dimension analysis to a broader range of circuit structures could provide a systematic basis for architecture design~\cite{PhysRevResearch.3.L032057}. Second, errors are inevitable in realistic quantum devices, making it important to assess the robustness of the architecture to gate errors and noise in the input data~\cite{Wang:2020yjh,Garcia-Martin:2023uuk}. Third, it would be interesting to investigate the computational capabilities of these circuits without training their quantum parameters, within the framework of quantum reservoir computing~\cite{Fujii:2016zzz,Mujal:2021crg,Xia:2021mex,Xia:2023wht,r8ww-qw7j,Chen:2020iry,Kobayashi:2024cgs,Ivaki:2025ora,Liu:2026etm,Wang:2026obv,Benavoli:2026scb,Martinez-Pena:2020rkm,Kornjaca:2024afu,Wang:2026gto,Ding:2026jrd}. The ability of contractive layers to control quantum information dynamics may also offer advantages in this setting~\cite{Martinez-Pena:2021qar}. Finally, incorporating mid-circuit measurements may enable more efficient generation of entanglement between qubits~\cite{Lu:2022jax,Baumer:2023vrf}. We leave these questions for future work.

  \vspace{5pt}
  \textit{Acknowledgements.}
  We thank Ning Sun and Yadong Wu for helpful discussions. Generative AI was used to assist with writing code and developing proofs in the Supplementary Material. We have verified the content and take responsibility for the correctness of this work. This project is supported by the Shanghai Rising-Star Program under grant number 24QA2700300, the NSFC under grant 12374477, the Quantum Science and Technology-National Science and Technology Major Project 2024ZD0300101, the Shanghai Qiguang Natural Science Development Foundation, and the Xuemin Institute of Advanced Studies at Fudan University.

\bibliography{ref.bib}

\end{document}


\title{Supplementary Material: Augmenting Quantum Neural Networks \\ with Contractive Layers}

  \author{Bolin Huang}
  \affiliation{State Key Laboratory of Surface Physics \& Department of Physics \& Center for Field Theory and Particle Physics, Fudan University, Shanghai, 200438, China}

  \author{Pengfei Zhang}
  \thanks{PengfeiZhang.physics@gmail.com}
  \affiliation{State Key Laboratory of Surface Physics \& Department of Physics \& Center for Field Theory and Particle Physics, Fudan University, Shanghai, 200438, China}
  \affiliation{Hefei National Laboratory, Hefei 230088, China}

  \date{\today}

  \begin{abstract}
  In this Supplemental Material, we provide details on (1) the generation of random target observables and input states; (2) the multilayer perceptron baseline; (3) the theoretical analysis of the local dimension; (4) image preprocessing and quantum state encoding; and (5) the classical readout for ten-class image recognition.
  \end{abstract}
    
  \maketitle

\section{Random generation of the target observable and input states}
\label{sec:random-generation}

For the observable learning task, we generated a single target observable on $n=5$ qubits and independently sampled a collection of input pure states. The Hilbert-space dimension was $D=2^n=32$. All circuit architectures and training runs used the same target observable and the same training and test sets. We also conducted tests using different random seeds, and the training results were similar to those reported in the paper.

\subsection{Target observable}
\label{subsec:target-observable}

We first sampled $D$ real eigenvalues independently from the uniform distribution on $[-2.5,2.5)$ and assembled them into a diagonal matrix,
\begin{equation}
    \Lambda = \operatorname{diag}(\lambda_1,\ldots,\lambda_{D}),
    \qquad
    \lambda_j \sim \operatorname{Uniform}[-2.5,2.5).
\end{equation}
We then generated a unitary $V$ using two brick-wall layers with open boundary conditions. Numbering the qubits from 1 to 5, each layer first applied two-qubit gates on $(1,2)$ and $(3,4)$, followed by gates on $(2,3)$ and $(4,5)$. Each gate was generated as
\begin{equation}
    G(\bm{\theta})
    = \exp\!\left(-i\sum_{P\in\mathcal{P}_2^\ast}\theta_P P\right),
    \qquad
    \mathcal{P}_2^\ast
    = \{I,X,Y,Z\}^{\otimes 2}\setminus\{I\otimes I\},
\end{equation}
where the 15 coefficients $\theta_P$ were sampled independently and uniformly from $[0,2\pi)$ for each gate. The target observable was
\begin{equation}
    \hat{O} = V\Lambda V^\dagger .
    \label{eq:target-observable}
\end{equation}
The resulting $\hat{O}$ was held fixed throughout the experiment.

\subsection{Input states and target expectation values}
\label{subsec:input-states}

For each input state, we independently sampled two real vectors $\boldsymbol{u},\boldsymbol{v}\in\mathbb{R}^{D}$ whose entries were independent standard normal random variables. We formed the complex vector with entries $z_j=u_j+i v_j$ and normalized it to obtain
\begin{equation}
    |\psi\rangle
    = \frac{\displaystyle\sum_{j=0}^{D-1}z_j|j\rangle}
           {\displaystyle\left(\sum_{j=0}^{D-1}|z_j|^2\right)^{1/2}} .
    \label{eq:input-state}
\end{equation}
We drew $4096$ training states followed by $1024$ test states from the same distribution.

For each state $|\psi_s\rangle$, the regression label was the exact expectation value of the fixed target observable,
\begin{equation}
    y(\psi_s) = \langle\psi_s|\hat{O}|\psi_s\rangle .
    \label{eq:target-label}
\end{equation}
We evaluated these labels directly from the state vectors and $\hat{O}$, without finite-shot sampling or added label noise. It is worth noting that all experiments were implemented on classical computers; we did not run these experiments on quantum hardware.

\section{Multilayer perceptron baseline}
\label{sec:mlp-baseline}
To provide a classical baseline for the observable learning task, we employed a multilayer perceptron (MLP) with two hidden layers. The network architecture was $64\rightarrow h\rightarrow h\rightarrow 1$, where the hidden dimension was varied over $h\in\{3,5,8,13,21,32\}$. For each input quantum state $|\psi\rangle$, the real and imaginary parts of its $32$ complex amplitudes were concatenated into a $64$-dimensional real vector, which was then used as the input to the MLP. Before being fed into the network, each input component was multiplied by a factor of $8$. Since the input states were sampled uniformly from the Haar distribution in a $32$-dimensional complex Hilbert space, each real component has variance $1/64$. This rescaling therefore ensures that each input component has unit variance, facilitating stable neural network training.

We used the Sigmoid Linear Unit (SiLU) ~\cite{Elfwing2017SigmoidWeightedLU}as the activation function in both hidden layers, defined as
\begin{equation}
    \operatorname{SiLU}(x)=\frac{x}{1+e^{-x}}.
\end{equation}
The SiLU activation provides a smooth, continuously differentiable nonlinearity, allowing gradients to vary smoothly during optimization. The final layer produces a single real-valued prediction of the target observable expectation value, without an additional activation function. The network was trained by minimizing the mean squared error (MSE) between the predicted and exact expectation values, using the Adam optimizer~\cite{Kingma2014AdamAM}. The same training and test datasets were used as in the quantum neural network experiments to ensure a consistent comparison.

\section{Local dimension}
\label{sec:pauli-rotation-dimension}

We consider an open chain of \(n\geq 4\) qubits and measure \(X_t\) on the central qubit \(t=\lfloor n/2\rfloor+1\), using one-based indexing. For even \(n\), reflection exchanges the two central qubits and preserves both rows of the brick-wall circuit, so either choice gives the same dimension. Each circuit unit consists of a brick-wall layer \(B\), followed by one or more trainable global Pauli-string rotations generated by \(X^{\otimes n}\), \(Y^{\otimes n}\), or \(Z^{\otimes n}\). For example, a global \(Z\) rotation followed by a global \(Y\) rotation is written as \(R(\alpha,\beta)=\exp(i\beta Y^{\otimes n})\exp(i\alpha Z^{\otimes n})\), with the rightmost factor acting first. We analyze both individual rotations and sequences of rotations about different global axes. The special case of a single global \(Z\) rotation,
\begin{equation}
  R(\alpha)
  =\hat{U}_{\text{CT}}(\alpha)
  =\exp\Big(i\alpha \prod_i\hat{Z}_i\Big),
  \end{equation}
is the contractive layer (CT) considered in the main text.

Building on Jacobian-based approaches to local expressivity in parameterized quantum circuits~\cite{Haferkamp2021LinearGO,Ohno:2026sve,Cha:2026byk,Funcke:2020vkw}, we determine the local dimension of the family of effective measurement operators generated by brick-wall layers and global Pauli-string rotations, accounting for the trainable readout amplitudes and biases. We obtain single-unit dimension counts classified by the number of visible terminal angles, establish their exactness analytically for \(n\geq9\), and certify the five-qubit results by exact finite-field calculations. We also derive depth-dependent upper bounds. For the five-qubit BW+CT architecture considered in the main text, the generic single-instance dimension is \(\kappa(5,L)=\min\{514,43+47(L-1)\}\), giving \(43,90,137,184\) at depths one through four. This is consistent with the numerical results in the paper. But such linear growth is not universal across system sizes and architectures. For multiple instances, we identify the tangent-space condition for additive dimension growth and show that a witnessed maximal rank is attained almost surely under continuous random parameter sampling. The common bias contributes only one dimension.

The derivation proceeds in four steps. “Local dimension of the effective measurement operator” relates the Pauli-coefficient Jacobian to operator tangents and identifies redundant directions through commutators, including the contributions of amplitude and bias. “One brick-wall layer followed by global Pauli rotations” derives single-unit ranks using analytical counts and exact finite-field checks. “Depth dependence and the five-qubit result” establishes depth-dependent bounds and their attainability for five qubits. “Dimension growth with the number of circuit instances” identifies when independently parameterized instances contribute additive dimensions. Readers interested only in the results may skip the detailed tangent-space and bridge-independence proofs and proceed to the rank formulas and discussions of depth and instance dependence. 

\subsection{Local dimension of the effective measurement operator}
\label{subsec:observable-tangent-dimension}

For a single circuit instance, let \(\boldsymbol{\vartheta}\) collect all circuit parameters, including the brick-wall parameters \(\bm{\theta}\) and the rotation-layer parameters, and let \(a,b\in\mathbb R\) be the trainable amplitude and bias. The full trainable parameter vector is \(\bm{\xi}=(\boldsymbol{\vartheta},a,b)\). Write \(I_n=I^{\otimes n}\) for the identity on the full Hilbert space. For an input state \(\rho\) of unit trace, the prediction is
\begin{align}
    \hat{y}(\rho)
    &=a\,\operatorname{Tr}\!\left[
        X_t U(\boldsymbol{\vartheta})\rho
        U(\boldsymbol{\vartheta})^\dagger\right]+b \notag\\
    &=\operatorname{Tr}\!\left(\rho \hat{M}(\bm{\xi})\right),
    \qquad
    \hat{M}(\bm{\xi})=aO(\boldsymbol{\vartheta})+bI_n,
    \qquad
    O(\boldsymbol{\vartheta})
       =U(\boldsymbol{\vartheta})^\dagger X_tU(\boldsymbol{\vartheta}).
    \label{eq:prediction-operator}
\end{align}
The second line follows from cyclicity of the trace and \(\operatorname{Tr}(\rho I_n)=1\). All reported dimensions of the effective measurement operator include both readout parameters and assume \(a\neq0\). For a pure input state \(\rho=|\psi\rangle\langle\psi|\), we write \(\hat{y}(\rho)=\hat{y}(\psi)\), consistently with the main text.

\paragraph{From parameter variations to prediction variations.}
Let \(\bm{\xi}=(\vartheta_1,\ldots,\vartheta_m,a,b)\). At a fixed parameter point, an infinitesimal displacement \(\delta\bm{\xi}\) produces the first-order change
\begin{equation}
    \delta \hat{M}
    =\sum_{j=1}^{m+2}\delta\xi_j
      \frac{\partial \hat{M}}{\partial\xi_j},
    \qquad
    \delta\hat{y}(\rho)=\operatorname{Tr}(\rho\,\delta \hat{M}).
    \label{eq:prediction-differential}
\end{equation}

It follows that the number of independent first-order changes at \(\bm{\xi}\) is
\begin{equation}
    d_{\mathrm{loc}}(\bm{\xi})
    =\dim_{\mathbb R}\operatorname{span}
       \left\{\frac{\partial \hat{M}}{\partial\xi_1},\ldots,
                   \frac{\partial \hat{M}}{\partial\xi_{m+2}}\right\}
    =\operatorname{rank}J(\bm{\xi})
    =(m+2)-\dim_{\mathbb R}\ker(\mathrm{d}\hat{M}_{\bm{\xi}}).
    \label{eq:local-independent-dimension}
\end{equation}
We denote the generic value of \(d_{\mathrm{loc}}(\bm{\xi})\) by \(\kappa\). At a specified parameter point, such as the end of training, \(d_{\mathrm{loc}}(\bm{\xi})\) in Eq.~\eqref{eq:local-independent-dimension} gives the local rank directly.

\paragraph{Equivalence to the Pauli-coefficient Jacobian.}
Let \(\mathcal P_n=\{I,X,Y,Z\}^{\otimes n}\) denote the Hermitian Pauli strings without phase factors. They satisfy
\begin{equation}
    2^{-n}\operatorname{Tr}(PP')=\delta_{P,P'}.
\end{equation}
Since \(X_t\) is traceless, \(\operatorname{Tr}(O)=0\) and \(\operatorname{Tr}(\hat{M})=2^n b\). The expansion therefore reads
\begin{equation}
    \hat{M}=\sum_{P\in\mathcal P_n} h_P P,\qquad
    h_{I_n}=2^{-n}\operatorname{Tr}(\hat{M})=b,\qquad
    h_P=a\,2^{-n}\operatorname{Tr}(PO)\quad(P\neq I_n).
\end{equation}
The identity coefficient depends only on \(b\); all nonidentity coefficients depend only on \(\boldsymbol{\vartheta}\) and \(a\). The coefficients are real because \(P\) and \(\hat{M}\) are Hermitian. Differentiating with respect to \(\xi_j\) gives
\begin{equation}
    \frac{\partial \hat{M}}{\partial\xi_j}
    =\sum_{P\in\mathcal P_n}J_{Pj}P,
    \qquad
    J_{Pj}=\frac{\partial h_P}{\partial\xi_j}
          =2^{-n}\operatorname{Tr}\!\left(
             P\frac{\partial \hat{M}}{\partial\xi_j}\right).
    \label{eq:pauli-jacobian-entries}
\end{equation}
Thus column \(j\) of \(J\) is the Pauli-coordinate vector of the operator \(\partial \hat{M}/\partial\xi_j\). For any real vector \(v\),
\begin{equation}
    \sum_j v_j\frac{\partial \hat{M}}{\partial\xi_j}
    =\sum_{P\in\mathcal P_n}(Jv)_P P.
\end{equation}
The Pauli strings are linearly independent, so the left-hand side vanishes if and only if \(Jv=0\). The two descriptions therefore have exactly the same null directions and the same rank:
\begin{equation}
    \operatorname{rank}_{\mathbb R}J
    =\dim_{\mathbb R}\operatorname{span}
       \left\{\frac{\partial \hat{M}}{\partial\xi_j}\right\}_j .
    \label{eq:pauli-tangent-rank}
\end{equation}
In particular, an analytical count can be performed directly with operators even when the numerical rank is computed in the Pauli basis. Ordering the identity coordinate and the bias parameter last gives the block form
\begin{equation}
    J=\begin{pmatrix}J_{aO}&0\\0&1\end{pmatrix},
    \qquad \operatorname{rank}J=\operatorname{rank}J_{aO}+1,
\end{equation}
where \(J_{aO}\) differentiates the nonidentity coefficients with respect to \((\boldsymbol{\vartheta},a)\). This identity also applies when evaluating the rank at saved training parameters.

\paragraph{How a brick-wall parameter gives a commutator.}
Consider one circuit unit \(U=RB\), with the brick-wall layer \(B\) executed first and the rotation layer \(R\) executed second. Hold the rotation angles fixed and define
\begin{equation}
    Q=R^\dagger X_tR,
    \qquad O=B^\dagger QB.
\end{equation}
For a brick-wall parameter \(\theta_j\), define its generator in the output frame of \(B\) by
\begin{equation}
    T_j=-i(\partial_jB)B^\dagger,
    \qquad \partial_j=\frac{\partial}{\partial\theta_j}.
\end{equation}
Differentiating \(BB^\dagger=I\) shows that \((\partial_jB)B^\dagger\) is anti-Hermitian, so \(T_j\) is Hermitian. Equivalently, \(\partial_jB=iT_jB\) and \(\partial_jB^\dagger=-iB^\dagger T_j\). Substitution into the product rule gives
\begin{align}
    \partial_jO
    &=(\partial_jB^\dagger)QB+B^\dagger Q(\partial_jB) \notag\\
    &=-iB^\dagger T_jQB+iB^\dagger QT_jB \notag\\
    &=iB^\dagger[Q,T_j]B.
    \label{eq:bw-observable-commutator}
\end{align}
If the parameter belongs to a gate \(V_j\) inside the brick-wall, write \(B=B_{>}V_jB_{<}\), where \(B_{>}\) contains the gates executed after \(V_j\). A gate-local variation \(\partial_jV_j=iG_jV_j\) then gives
\begin{equation}
    T_j=B_{>}G_jB_{>}^\dagger.
\end{equation}

The same calculation detects dependencies among several parameters. For real coefficients \(v_j\), let \(T(v)=\sum_jv_jT_j\). Then
\begin{equation}
    \sum_jv_j\,\partial_jO
       =iB^\dagger[Q,T(v)]B.
\end{equation}
Since conjugation by \(B\) is invertible, the combination on the left vanishes precisely when \([Q,T(v)]=0\). Thus two nonzero parameter derivatives can be dependent even when neither derivative vanishes individually. Define the real brick-wall generator space and the Hermitian commutant of \(Q\) by
\begin{equation}
    \mathcal G_B=\operatorname{span}_{\mathbb R}\{T_j\}_j,
    \qquad
    \mathcal C(Q)=\{T=T^\dagger:[Q,T]=0\}.
\end{equation}
Applying rank--nullity to the map \(T\mapsto i[Q,T]\) restricted to \(\mathcal G_B\) gives
\begin{equation}
    \dim_{\mathbb R}\operatorname{span}
       \{\partial_jO:\theta_j\text{ belongs to } B\}
    =\dim_{\mathbb R}\mathcal G_B
     -\dim_{\mathbb R}\bigl(\mathcal G_B\cap\mathcal C(Q)\bigr).
    \label{eq:brick-wall-visible-dimension}
\end{equation}
This is the subtraction used in the analytical dimension count. Taking the span in \(\mathcal G_B\) first accounts for different brick-wall parameters that generate the same unitary variation. The intersection with \(\mathcal C(Q)\) then accounts for distinct unitary variations that leave the measured observable unchanged. If a parameter \(\phi\) belongs to \(R\), its derivative is instead \(B^\dagger(\partial_\phi Q)B\). These rotation derivatives must be included in the same span when counting the complete circuit.

\paragraph{Independent contributions of amplitude and bias.}
Unitary conjugation preserves \(X_t^2=I_n\) and \(\operatorname{Tr}(X_t)=0\). Differentiating \(O^2=I_n\) and \(\operatorname{Tr}(O)=0\) with respect to a circuit angle gives
\begin{equation}
    O\,\partial_jO+(\partial_jO)\,O=0,\qquad
    \operatorname{Tr}(O\,\partial_jO)=0,\qquad
    \operatorname{Tr}(\partial_jO)=0,
\end{equation}
where the middle identity follows by taking the trace of the first. The derivatives of the effective measurement operator are
\begin{equation}
    \partial_j \hat{M}=a\,\partial_jO,\qquad
    \partial_a \hat{M}=O,\qquad \partial_b \hat{M}=I_n.
\end{equation}
The amplitude direction \(O\) is orthogonal to every circuit derivative in the Hilbert--Schmidt inner product. The bias direction \(I_n\) is orthogonal to \(O\) and to every circuit derivative because they are traceless. Both readout directions have squared norm \(2^n\). Since \(a\neq0\), the circuit derivatives of \(\hat{M}\) span the same space as those of \(O\). Hence
\begin{equation}
    d_{\mathrm{loc}}(\bm{\xi})
    =\dim_{\mathbb R}\operatorname{span}
       \{\partial_{\vartheta_j}(U^\dagger X_tU)\}_j
      +\underbrace{1}_{\text{amplitude}}
      +\underbrace{1}_{\text{bias}}.
    \label{eq:amplitude-plus-circuit-dimension}
\end{equation}
The same identity holds for the generic dimension \(\kappa\). Each readout parameter contributes once to the effective measurement operator for the entire circuit.

\subsection{One brick-wall layer followed by global Pauli rotations}
\label{subsec:single-global-pauli-block}

The first brick-wall row contains independent \(SU(4)\) gates on \((1,2),(3,4),\ldots\), and the second row contains independent \(SU(4)\) gates on \((2,3),(4,5),\ldots\). There are \(n-1\) two-qubit gates. On each of the \(n-2\) wires shared by the rows, a one-qubit \(SU(2)\) basis change can be inserted in the first row and canceled in the second. The brick-wall unitary therefore has
\begin{equation}
    D_B=15(n-1)-3(n-2)=12n-9
    \label{eq:brick-wall-unitary-dimension}
\end{equation}
independent directions. Equivalently, the tangent space can be decomposed into block-local and inter-block directions, giving \(D_B=15p+3u+9\lfloor n/2\rfloor\), where \(p=\lfloor(n-1)/2\rfloor\) and \(u=n-2p\). This decomposition will be useful for identifying measurement-invisible directions below.

Let \(\overline P=P^{\otimes n}\) and \(R_P(\phi)=\exp(i\phi\overline P)\). Denote by \(k\) the number of independent terminal angles visible in \(R^\dagger X_tR\). The parity of \(n\) determines \(k\), since
\begin{equation}
    \overline X\,\overline Y
    =(-1)^n\overline Y\,\overline X .
\end{equation}
For odd \(n\), the global strings generate an \(SU(2)\) action. Writing \(X_t=\overline X S\), with \(S=\prod_{j\neq t}X_j\), shows that \(S\) commutes with all three global axes. The terminal orbit is therefore the orbit of a single axis on a sphere and has dimension at most two. A single \(X\) rotation gives \(k=0\), whereas a single \(Y\) or \(Z\) rotation gives \(k=1\). For two distinct axes, \(k=1\) if the last executed rotation is \(X\), and \(k=2\) otherwise. Any permutation of all three axes gives \(k=2\). For even \(n\), the global generators commute, and the global \(X\) rotation drops out of the terminal observable. Here \(k\) is the number of distinct axes from \(\{Y,Z\}\) present in the layer.

To count the remaining brick-wall directions, expand \(Q=\sum_s c_sQ_s\) in its visible Pauli strings. On each output block, count the traceless Hermitian generators that commute with the local restrictions of every \(Q_s\) with \(c_s\neq0\). For a central measurement, exactly one two-qubit block contains \(t\). The dimensions of these block-local commuting spaces are
\begin{equation}
\begin{array}{c|ccc}
 k & \text{singleton}
   & \text{two-qubit block containing }t
   & \text{other two-qubit block}\\ \hline
 1 & 1 & 3 & 7\\
 2 & 0 & 1 & 3
\end{array}
\label{eq:global-block-commutants}
\end{equation}
For example, at \(k=1\) an unmeasured two-qubit block sees a single nontrivial Pauli product such as \(Y\otimes Y\). Its two eigenspaces both have dimension two, so its traceless commutant has dimension \(2^2+2^2-1=7\). At \(k=2\), the same block must preserve both \(Y\otimes Y\) and \(Z\otimes Z\). Their four joint eigenspaces are one-dimensional, leaving \(4-1=3\) traceless directions. On the measured block, with the measured qubit written second, representative restrictions are \(IX,YZ,ZY\). The traceless common commutant of the first two has dimension three; imposing the third leaves dimension one. For even \(n\) with \(k=2\), the additional terminal term restricts to \(XI\) on the measured block and \(XX\) on an unmeasured two-qubit block. These impose no further commuting constraints because \(XI=(IX)(YZ)(ZY)\) and \(XX=-(YY)(ZZ)\). On a singleton, commuting with both \(Y\) and \(Z\) already leaves no nonzero traceless direction.

Let \(\kappa_{1,k}\) denote the single-unit dimension of the effective measurement operator, including the amplitude and bias. Subtracting the commuting directions in Eq.~\eqref{eq:global-block-commutants} from \(D_B+k+2\) gives
\begin{equation}
\begin{aligned}
    \kappa_{1,1}
    &=D_B+3-\bigl[u+3+7(p-1)\bigr]
      =12n-2-u-7p,\\
    \kappa_{1,2}
    &=D_B+4-\bigl[1+3(p-1)\bigr]
      =12n-3-3p.
\end{aligned}
\label{eq:single-unit-global-ranks}
\end{equation}
The case \(k=0\) has a different derivation. The terminal rotation commutes with \(X_t\), leaving only three gates in the backward light cone of the measured observable: the second-row gate containing \(t\) and its two first-row neighbors. The measured gate contributes \(16-2(2^2)=8\) orbit directions. Each neighboring gate contributes nine additional directions, distinguished by a nonidentity Pauli on its outer wire. At a generic measured gate the three local Pauli generators on either inner wire have independent commutators with the transported observable. Including both readout directions gives
\begin{equation}
    \kappa_{1,0}=8+9+9+\underbrace{1}_{a}+\underbrace{1}_{b}=28.
\end{equation}

\paragraph{Why different terminal Pauli strings cannot cancel.}
For \(k=1,2\), the count in Eq.~\eqref{eq:single-unit-global-ranks} subtracts the known block-local commuting directions. Equality requires showing that these exhaust the kernel and that the surviving brick-wall directions are independent of the terminal-angle and readout directions. Work in the output frame of the brick-wall, where a brick-wall variation is represented by \(i[Q,T]\), with \(Q=\sum_s c_s Q_s\).

Expand \(T\) in Pauli strings \(A\). A generator of a second-row gate acts on one output block. A generator of a first-row gate is conjugated by at most two second-row gates, one meeting each end of its edge. Hence each Pauli component \(A\) occupies at most four qubits, even when the sum \(T\) extends across the chain. For each terminal string \(Q_s\), the commutator \([Q_s,A]\) is either zero or proportional to \(AQ_s\). Suppose components \(A,A'\) from two distinct terminal strings contributed to the same Pauli coefficient. Up to a Pauli phase, this would require
\[
    AQ_s \propto A'Q_{s'}
    \quad\Longrightarrow\quad
    (A')^\dagger A \propto Q_{s'}Q_s^\dagger .
\]
The left-hand side occupies at most eight qubits. To see the support on the right, write \(Q_s\propto G_sX_t\), where \(G_s\) is the identity or one of the three global Pauli strings. The factors \(X_t\) cancel, and for \(s\neq s'\) the product \(G_{s'}G_s^\dagger\) is nonidentity on every qubit. For \(n\geq9\) the two supports therefore cannot agree. In particular,
\begin{equation}
    [Q,T]=0
    \quad\Longleftrightarrow\quad
    [Q_s,T]=0\quad\text{for every }s\text{ with }c_s\neq0 
\end{equation}
This is an intersection of kernels: the same \(T\) must commute with every visible term. For a fixed \(Q_s\), multiplication by \(Q_s\) is one-to-one on Pauli strings up to phase. We refer to a first-row two-qubit gate connecting two distinct second-row output blocks as a bridge. Consequently, components supported within a single output block cannot cancel components acting nontrivially on both blocks connected by a bridge. Similarly, different bridges connect distinct pairs of output blocks, so their contributions cannot cancel one another. It therefore remains to examine the kernel associated with each bridge separately.

\paragraph{Why each first-row bridge contributes nine directions.}
Choose a visible \(Q_s\) with full support. Such a term exists at generic angles for \(k\geq1\): at least one term proportional to \(\overline YX_t\) or \(\overline ZX_t\) has nonzero coefficient. A bridge joins two adjacent output blocks, at least one of which is a two-qubit block. Call that block the left factor; the other factor may be a singleton at the chain boundary. Move to the frame immediately before the output-block gates. In this frame the bridge space is \(V_L\otimes V_R\), where each \(V\) is the three-dimensional real span of \(X,Y,Z\) on the wire touched by the bridge. Write \(P_L,P_R\) for the restrictions of the chosen terminal string to the two blocks, conjugated into the same frame. Its factors on all other blocks are invertible and can be dropped from the commutator test. A bridge variation \(K\) is therefore invisible to this term precisely when \([P_L\otimes P_R,K]=0\).

For a Hermitian involution \(P\), define
\begin{equation} 
    \Pi_\pm^P(A)=\tfrac12(A\pm PAP).
\end{equation}
These projections extract the parts of \(A\) that commute and anticommute with \(P\), respectively. On the two-qubit end, both projections are injective on \(V_L\) at a generic output gate. To verify this, order the local factors so that the bridge touches the first wire. The nonidentity two-qubit Pauli on this block has eigenvalues \(+1,-1\), each twice. An allowed \(SU(4)\) conjugation can therefore place it at
\begin{equation} 
    P_*=\frac{IZ+ZI+XY-YX}{2}
       =D(I\otimes Z)D^\dagger,\qquad
    D=\frac{I+i\,\mathrm{SWAP}}{\sqrt{2}},
\end{equation}
where juxtaposition denotes a tensor product and an overall phase of \(D\) has no effect. Take \(A=C\otimes I\) with \(C=v_xX+v_yY+v_zZ\). If \([A,P_*]=0\), the terms \(ZI,XY,YX\) have distinct right factors, so \(C\) must commute separately with \(Z,X,Y\); hence \(C=0\). If \(\{A,P_*\}=0\), the component with right factor \(Z\) comes solely from \(IZ\) and is \(C\otimes Z\); again \(C=0\). Thus both projections are injective on \(V_L\) at the constructed gate, so each has a nonzero \(3\times3\) minor. Each such minor is a real-analytic function of the gate parameters. Its nonzero value at this particular gate proves that it is not identically zero, and its zero set therefore has measure zero on the connected group \(SU(4)\). Failure of injectivity can occur only within this zero set. Since there are finitely many projections and bridges, the union of their exceptional sets also has measure zero. Hence all the required injectivity conditions hold simultaneously at generic gates.

Now suppose \(K\in V_L\otimes V_R\) commutes with \(P_L\otimes P_R\). Conjugation by this product reverses the sign of the two components with opposite left and right parities. These components belong to distinct subspaces, so both vanish:
\begin{equation}
    (\Pi_+^{P_L}\otimes\Pi_-^{P_R})K=0,\qquad
    (\Pi_-^{P_L}\otimes\Pi_+^{P_R})K=0 .
\end{equation}
Apply injectivity on the left factor to the first equation. It gives \((\mathrm{id}\otimes\Pi_-^{P_R})K=0\). Here the identity paired with the injective left map acts on the full right-block operator space, which contains the projected right factors. The second equation gives \((\mathrm{id}\otimes\Pi_+^{P_R})K=0\) by the same argument. Adding the equations and using \(\Pi_+^{P_R}+\Pi_-^{P_R}=\mathrm{id}\) yields \(K=0\). The bridge commutator map is therefore injective on a space of dimension \(3\times3=9\), including at a boundary bridge.

\paragraph{Terminal angles, amplitude, and bias.}
At generic angles, differentiating \(Q=\sum_s c_sQ_s\) changes only the coefficients. The \(k\) independent terminal-angle derivatives and the amplitude direction \(Q\) thus lie in the span of the terminal strings. A nonzero brick-wall commutator has no component in this span. Indeed, \(AQ_s\propto Q_{s'}\) would require \(A\) to be a full-support string for \(s'\neq s\), while \(s'=s\) would require \(A\) to be the identity, whose commutator vanishes. The \(k\) terminal directions are orthogonal to \(Q\) because \(Q^2=I_n\). The bias derivative \(I_n\) is orthogonal to all these traceless directions. This proves that the two readout parameters contribute one dimension each and that the upper counts in Eq.~\eqref{eq:single-unit-global-ranks} are attained for \(n\geq9\).

For \(n=5\), we verify the ranks exactly by constructing the observable Jacobian at circuit points with rational real and imaginary matrix entries and performing Gaussian elimination modulo the prime \(1009\). A nonzero minor modulo \(1009\) certifies a nonzero minor of the original Jacobian, providing a rigorous rank lower bound~\cite{Cha:2026byk}. When this matches the analytical upper bound, the generic rank is determined exactly. The amplitude derivative is included as the column \(O\), and the independent bias derivative \(I_n\) contributes one additional dimension.

\subsection{Depth dependence and the five-qubit result}
\label{subsec:global-pauli-depth}

Adding a circuit unit introduces new tangent directions, some of which may overlap with those already generated. If \(V\) is the old circuit tangent space and \(W\) is the space contributed by the new unit in the same frame, the dimension increment is
\begin{equation}
    \Delta d_{\mathrm{loc}}=\dim W-\dim(V\cap W).
    \label{eq:exact-increment}
\end{equation}
For one global axis, \(n\) local same-axis rotations can be absorbed across the interface, giving the increment bound \(s_1=D_B+1-n=11n-8\). For two or three distinct axes, parameter counting gives \(s_2=D_B+2\) and \(s_3=D_B+3\). Combining these bounds with the observable-orbit dimension, and counting amplitude and bias once, yields
\begin{equation}
    \kappa(n,L)\leq
    \min\!\left\{2^{2n-1}+2,\ \kappa_{1,k}+(L-1)s_g\right\}.
    \label{eq:global-depth-bound}
\end{equation}
Here \(g\in\{1,2,3\}\) denotes the number of distinct global rotation axes in each circuit unit.

For five qubits with central measurement \(X_3\), exact finite-field Jacobian witnesses attain this bound through first saturation. Adding identity units then establishes equality at all larger depths. For example, BW followed by a global \(Z\) rotation gives dimensions \(43,90,137,184\) at depths one through four. Equality depends on the circuit size and architecture. For nine qubits with central measurement \(X_5\), two BW+\(X\) units have dimension \(114\), below the bound of \(119\), because an input-side gate outside the light cone of the output-side measured observable contributes five additional redundant directions.

\subsection{Dimension growth with the number of circuit instances}
\label{subsec:independent-circuit-instances}

Fix the circuit depth \(L\) and consider \(R\) independently parameterized instances acting on the same input state. With the linear classical readout of the main text (including trainable biases), their predictions add. The effective measurement operator is
\begin{equation}
    \hat{M}(\bm{\xi})=bI_n+\sum_{r=1}^{R}a^r O_r,\qquad
    O_r=\hat{U}^{r}(\boldsymbol{\vartheta}_r)^\dagger X_t
        \hat{U}^{r}(\boldsymbol{\vartheta}_r).
    \label{eq:sum-instance-observable}
\end{equation}
If the implementation trains a separate bias \(b^r\) for each instance, then \(b=\sum_r b^r\). Every bias derivative is the same operator \(I_n\), so all biases together contribute exactly one dimension.

\paragraph{When do the dimensions add?}
Varying a parameter of instance \(r\) changes only its own summand:
\[
    \partial_{\vartheta_{rj}}\hat{M}=a^r\,\partial_{\vartheta_{rj}}O_r,
    \qquad \partial_{a^r}\hat{M}=O_r .
\]
Thus adding an instance preserves all existing Jacobian columns and appends the columns of the new instance. Let
\[
    V_r=\operatorname{span}_{\mathbb R}
        \{O_r,\ a^r\partial_{\vartheta_{rj}}O_r:\ 1\leq j\leq m_r\},
    \qquad S_R=V_1+\cdots+V_R ,
\]
where \(m_r\) is the number of circuit parameters of instance \(r\). Each \(V_r\) lies in the traceless Hermitian operator space. The bias direction is orthogonal to that entire space, giving the exact identities
\begin{align}
    d_{\mathrm{loc}}^{(R)}(\bm{\xi}_R) &= 1+\dim S_R,\\
    d_{\mathrm{loc}}^{(R+1)}(\bm{\xi}_{R+1})-d_{\mathrm{loc}}^{(R)}(\bm{\xi}_R)
       &=\dim V_{R+1}-\dim(V_{R+1}\cap S_R).
    \label{eq:instance-dimension-increment}
\end{align}
Here \(\bm{\xi}_R\) collects all circuit and readout parameters for the \(R\)-instance model, and the old instances are evaluated at the same parameter values on both sides. The increment is the full dimension of the new instance precisely when its tangent space intersects the combined old space only at zero.

Let \(q_L\) be the generic single-instance dimension including its amplitude and excluding the common bias, so that \(\kappa_1=1+q_L\) at a generic point. Since the traceless Hermitian space has real dimension \(4^n-1\), identical architectures satisfy
\begin{equation}
    \kappa_R\leq 1+\min\{R q_L,\ 4^n-1\}.
    \label{eq:instance-dimension-bound}
\end{equation}

\paragraph{Why random parameter points attain the additive bound.}
Fix \(R,L\), and set \(m=\min\{Rq_L,4^n-1\}\). Let \(J_R^0\) be the Jacobian of the traceless part \(\sum_r a^rO_r\) in Pauli coordinates, with the amplitude derivatives included. It suffices to exhibit one parameter tuple at which \(J_R^0\) has rank \(m\). At that tuple, some \(m\times m\) minor has nonzero determinant. Keep those rows and columns fixed and denote their determinant by \(\Delta(\boldsymbol{\vartheta}_1,\ldots,\boldsymbol{\vartheta}_R)\). We may take all \(a^r=1\) when forming the minor: any other nonzero \(a^r\) merely rescales the circuit-derivative columns of instance \(r\), leaving its amplitude column and the rank unchanged.

The gate matrices depend real-analytically on their parameters, so \(\Delta\) is real-analytic on the connected circuit parameter domain. Since it is nonzero at the exhibited tuple, its zero set has measure zero. In particular, independent draws of the circuit parameters from distributions with densities, such as the Gaussian draws used here, satisfy
\begin{equation}
    \Pr\!\left[\operatorname{rank}J_R^0<m\right]
    \leq \Pr\!\left[\Delta=0\right]=0.
\end{equation}
Together with Eq.~\eqref{eq:instance-dimension-bound}, this gives \(\kappa_R=1+m\) almost surely~\cite{Haferkamp2021LinearGO}. When \(Rq_L\leq4^n-1\), the instance tangent spaces therefore form a direct sum almost surely. Above that threshold, they span the full traceless space. Both conclusions use the existence of a rank-\(m\) tuple for the chosen architecture.

At depth \(L=1\), the five-qubit BW+CT circuit has \(48\) Pauli coefficients that vanish identically for any number of instances \(R\), corresponding to the strings \(P=I_{12}\otimes P_{34}\otimes P_5\) with \(P_5\in\{X,Y,Z\}\). For each instance, \(\operatorname{Tr}(P O_r)=0\) because the terminal \(X_3\) term remains proportional to the identity on qubit \(5\) after conjugation by the brick-wall layer and thus contributes a factor \(\operatorname{Tr}(P_5)=0\), whereas the terminal \(Z_1Z_2Y_3Z_4Z_5\) term has zero partial trace over qubits \(1\) and \(2\) after the same conjugation. For \(L>1\), the local dimension eventually saturates at 1024.

\section{Image preprocessing and quantum state encoding}
\label{sec:image-encoding}
For the image recognition task, we used the Street View House Numbers (SVHN) dataset~\cite{Netzer2011ReadingDI}, where each input image has a resolution of $32\times32$ pixels with three RGB color channels. To encode each image into a five-qubit quantum state, we first converted the RGB image into grayscale using the fixed weighted combination
\begin{equation}
    g_{uv}=0.2989R_{uv}+0.5870G_{uv}+0.1140B_{uv},
\end{equation}
where $R_{uv}$, $G_{uv}$, and $B_{uv}$ denote the three color channels at pixel position $(u,v)$. We then applied non-overlapping $4\times4$ average pooling with a stride of $4$, reducing the image from $32\times32$ to $8\times8$ pixels. The pooled pixel values were calculated as
\begin{equation}
    v_{ab}=\frac{1}{16}\sum_{i=0}^{3}\sum_{j=0}^{3}
    g_{4a+i,4b+j},
    \qquad a,b=0,\ldots,7.
\end{equation}
The resulting $8\times8$ grayscale image was flattened in row-major order into a $64$-dimensional real vector $\boldsymbol{v}=(v_0,\ldots,v_{63})$. We then paired consecutive elements to construct $32$ complex amplitudes,
\begin{equation}
    a_j=v_{2j}+iv_{2j+1},
    \qquad j=0,\ldots,31.
\end{equation}
Finally, the complex vector was normalized to unit norm to obtain the five-qubit input state,
\begin{equation}
    |\psi\rangle
    =\frac{\displaystyle\sum_{j=0}^{31}a_j|j\rangle}
    {\displaystyle\left(\sum_{j=0}^{31}|a_j|^2\right)^{1/2}}.
    \label{eq:image-encoding}
\end{equation}
Since the resulting state contains $32=2^5$ complex amplitudes, it can be represented using five qubits. All preprocessing operations were fixed and involved no trainable parameters. The same encoding procedure was applied to both training and test images. The normalized state vectors were directly used as inputs to the classically simulated quantum circuits, without implementing physical quantum state preparation.

\section{Classical readout for ten-class image recognition}
\label{sec:recognition-readout}

For an input state $|\psi\rangle$, each of the $R$ circuit instances produces one real-valued feature $m^r(\psi)\in[-1,1]$, the expectation value of the Pauli-$X$ observable on the central qubit. To convert these $R$ features into class probabilities, we assign a separate fifth-degree polynomial to each instance $r$ and each digit class $k\in\{0,\ldots,9\}$. The score (logit) for class $k$ is
\begin{equation}
    z_k(\psi)
    = \sum_{r=1}^{R}\sum_{\nu=0}^{5}
      \beta_{rk\nu}\bigl[m^r(\psi)\bigr]^\nu ,
    \label{eq:class-logits}
\end{equation}
where $\beta_{rk\nu}$ is the trainable coefficient of order $\nu$ for instance $r$ and class $k$. Each polynomial acts on a single measurement expectation value, and the contributions from all instances are added directly. The $\nu=0$ terms contribute a class bias
$b_k=\sum_{r=1}^{R}\beta_{rk0}$. 

We then apply a softmax over the ten class scores,
\begin{equation}
    p_k(\psi)
    = \frac{\exp[z_k(\psi)]}
           {\displaystyle\sum_{j=0}^{9}\exp[z_j(\psi)]},
    \qquad k=0,\ldots,9 .
    \label{eq:class-probabilities}
\end{equation}
Thus $p_k(\psi)$ is the predicted probability that the input image depicts digit $k$, with $\sum_{k=0}^{9}p_k(\psi)=1$. For a training example \(\psi_s\) with true label \(y(\psi_s)\), the cross entropy loss is \(-\log p_{y(\psi_s)}(\psi_s)\). The polynomial coefficients and the quantum circuit parameters are optimized jointly. At inference, the predicted digit is the class with the largest score, $\hat{y}(\psi)=\arg\max_k z_k(\psi)$.

\bibliography{ref.bib}